# Magnetic diffuse scattering of the $S$ = 5/2 fcc antiferromagnets Ba$_2$MnTeO$_6$ and Ba$_2$MnWO$_6$

Otto H. J. Mustonen,[1,2]* Charlotte E. Pughe,[3] Lucile Mangin-Thro,[4] Robert Tarver,[2] Heather M. Mutch,[3] Helen C. Walker,[5] Edmund J. Cussen[6,3]*

[1]Department of Chemistry and Materials Science, Aalto University, FI-00076 Aalto, Finland

[2]School of Chemistry, University of Birmingham, Birmingham B15 2TT, United Kingdom

[3]Department of Material Science and Engineering, University of Sheffield, Mappin Street, Sheffield S1 3JD, United Kingdom

[4]Institut Laue-Langevin, 71 avenue des martyrs, 38000 Grenoble, France

[5]ISIS Pulsed Neutron and Muon Source, STFC Rutherford Appleton Laboratory, Harwell Campus, Didcot OX11 0QX, United Kingdom

[6]TU Dublin - School of Chemical and BioPharmaceutical Sciences, Grangegorman, D07 ADY7, Ireland

* Corresponding authors:

Otto H. J. Mustonen

otto.mustonen@aalto.fi

Edmund J. Cussen

e.j.cussen@sheffield.ac.uk

edmund.cussen@tudublin.ie

We have investigated the magnetic diffuse scattering of isostructural $S$ = 5/2 fcc antiferromagnets Ba$_2$MnTeO$_6$ and Ba$_2$MnWO$_6$ using polarized neutrons. Both materials display short-range correlated magnetism above their respective magnetic ordering temperatures of 20 K and 8 K. The spin correlations were analysed using a Reverse Monte Carlo approach. For Ba$_2$MnTeO$_6$, we find antiferromagnetic nearest-neighbor correlations along with ferromagnetic next-nearest neighbor correlations directly linked to the Type I order below $T_N$. For Ba$_2$MnWO$_6$, both the nearest-neighbor and next-nearest-neighbor spin correlations are antiferromagnetic in the paramagnetic state. The short-range spin correlations persist up to $T$ = 7$T_N$. The magnetic diffuse scattering was also fitted using Onsager reaction-field theory allowing us to evaluate the magnetic interactions in these materials. We obtained $J_1$ = -3.25(3) K and $J_2$ = 0.41(2) K for Ba$_2$MnTeO$_6$ and $J_1$ = -1.08(1) K and $J_2$ = -0.88(1) K for Ba$_2$MnWO$_6$. These interactions are comparable to previous results from inelastic neutron scattering experiments below $T_N$, which highlights the potential of the Onsager approach for the analysis of magnetic interactions.

# I. INTRODUCTION

Magnetic frustration occurs when the magnetic interactions in a material cannot be fully satisfied simultaneously [1]. It can arise from geometric frustration, often linked to triangular structural motifs, or competition between different interactions. Magnetic frustration can stabilize unusual magnetic states such as quantum spin liquids [2,3], quantum spin ice [4], valence bond solids [5] or valence bond glasses [6,7]. The face-centered cubic (fcc) lattice is an example of a geometrically frustrated lattice, where the nearest-neighbor interactions are frustrated and cannot be simultaneously satisfied. Magnetism of many fcc antiferromagnets can be described using the fcc $J_1$-$J_2$ Heisenberg model:

$$H = -J_1 \sum_{(i,j)} \mathbf{S}_i \cdot \mathbf{S}_j - J_2 \sum_{\langle i,j \rangle} \mathbf{S}_i \cdot \mathbf{S}_j \quad (1)$$

where $J_1$ is the nearest-neighbor exchange, $J_2$ is the next-nearest-neighbor exchange, $\mathbf{S}_i$ is the spin at site $i$, and the sums are taken over each bond. In this article, we make the sign choice that positive (negative) interactions are (anti)ferromagnetic. This model can result in three different antiferromagnetic structures – Type I, Type II and Type III – depending on the relative strength of the $J_1$ and $J_2$ interactions [8,9].

The archetypical fcc antiferromagnet is MnO. It crystallizes in the NaCl structure with an fcc lattice of $S = 5/2$ $Mn^{2+}$ cations and has a magnetic transition at $T_N$ = 118 K. Shull and Smart [10] provided the first direct evidence for antiferromagnetism with neutron diffraction experiments on MnO. They observed magnetic Bragg peaks below $T_N$ indicating doubling of the crystallographic cell along $a$, $b$ and $c$ directions. This corresponds to the Type II antiferromagnetic structure [10,11]. The magnetic transition is accompanied by a small structural distortion, which reduces the frustration of the nearest-neighbor interactions. Significant magnetic diffuse scattering persists in the paramagnetic state above $T_N$. This arises from short-range spin correlations in the form of local clusters [12,13]. The spin correlations above $T_N$ differ from the correlations in the ordered state: the frustration related to the nearest-neighbor spins is lifted in the paramagnetic state [13,14]. Moreover, the short-range spin correlations are observed up to a very high temperature of 1100 K [15].

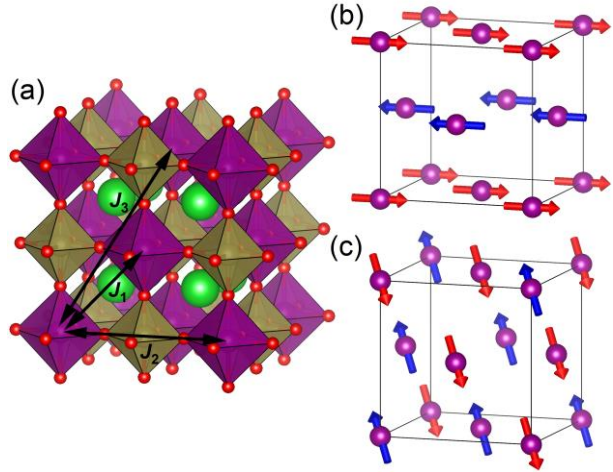

Figure 1. (a) The B-site ordered double perovskite structure of $Ba_2MnTeO_6$ and $Ba_2MnWO_6$. These materials crystallize in the cubic space group $Fm\bar{3}m$. The $Mn^{2+}$ cations on the B'-site (purple) form an fcc lattice. The magnetic structure depends on the non-magnetic B''-cations (yellow), which link the B'-sites via extended superexchange interactions. (b) The Type I magnetic structure of $Ba_2MnTeO_6$ with **k** = (0, 0, 1). (c) The Type II magnetic structure of $Ba_2MnWO_6$ with **k** = (1/2, 1/2, 1/2).

An fcc lattice of magnetic cations is also formed in the B-site ordered double perovskites $A_2B'B''O_6$ with NaCl ordering of the octahedral B' and B'' sites [16]. Cubic $Mn^{2+}$ double perovskites $Ba_2MnTeO_6$ and $Ba_2MnWO_6$ are S = 5/2 fcc antiferromagnets analogous to MnO [17,18]. They crystallise in the space group $Fm\bar{3}m$ with complete ordering of $Mn^{2+}$ on the B'-site and $Te^{6+}$ or $W^{6+}$ on the B''-site as shown in Figure 1a [17–20]. Magnetism in these materials is well described by the $J_1$-$J_2$ model with two interactions: nearest-neighbor $J_1$ from corner to face center ($r_1 \approx 5.8$ Å) and next-nearest-neighbor $J_2$ from corner to corner ($r_2 \approx 8.2$ Å) [17,18,21]. These are Mn – O – Te/W – O – Mn extended superexchange interactions mediated by the non-magnetic $d^{10}$ $Te^{6+}$ or $d^0$ $W^{6+}$ cations on the B''-site. Differences in orbital hybridization of $d^{10}$ and $d^0$ cations with O 2p lead to different dominant interactions and magnetic ground states for $Ba_2MnTeO_6$ and $Ba_2MnWO_6$ [17,22–25]. $Ba_2MnTeO_6$ magnetically orders at $T_N$ = 20 K into a Type I antiferromagnetic structure with **k** = (0, 0, 1) [17,26], see Figure 1b. The $Mn^{2+}$ spins align ferromagnetically in the (001) plane and adjacent planes couple antiferromagnetically. In contrast, $Ba_2MnWO_6$ orders at $T_N$ = 8 K into the Type II antiferromagnetic structure with **k** = (1/2, 1/2, 1/2) [18–20], see Figure 1c. The $Mn^{2+}$ spins form ferromagnetic (111) layers with antiferromagnetic coupling between adjacent layers. A variation of the Type II structure is also observed in $Ba_2MnMoO_6$, where $4d^0$ $Mo^{6+}$ is located on the linking B''-site analogous to $5d^0$ $W^{6+}$ in $Ba_2MnWO_6$ [27].

It should be noted that $Ba_2MnTeO_6$ is sometimes incorrectly described as a triangular lattice antiferromagnet in the literature [26,28] based on refinements in the lower symmetry subgroup

$R\bar{3}m$ [29]. This is a misunderstanding of the crystal structure, which actually hosts four such "triangular lattices" [17,30]. These correspond to the four threefold rotation axes along body diagonals that define cubic lattice symmetry [31].

Given the similarities with MnO, does short-range correlated magnetism persist above $T_N$ also in Ba$_2$MnTeO$_6$ and Ba$_2$MnWO$_6$? A number of previous studies suggest this is the case, but the spin correlations have never been characterised. Both Ba$_2$MnTeO$_6$ and Ba$_2$MnWO$_6$ are moderately frustrated with $f = |\theta_{CW}|/T_N \approx 8$ [17,18]. For Ba$_2$MnTeO$_6$, magnetic diffuse scattering was observed above $T_N$ using neutron diffraction similar to MnO [17]. Magnetic excitations were observed in the inelastic neutron scattering up to 109 K [17] likely arising from short-range correlated magnetism. Moreover, transverse-field muon spin rotation and relaxation measurements suggest short-range order starts to develop below 35 K [28]. Similarly, inelastic neutron scattering experiments on Ba$_2$MnWO$_6$ revealed magnetic excitations up to at least 40 K, and transverse-field muon experiments suggest the formation of a short-range correlated state below 30 K [18].

Correlated spins in the paramagnetic phase give rise to magnetic diffuse scattering [12]. Polarized neutron scattering is an ideal method for investigating spin correlations, because it allows for the separation of the magnetic and nuclear diffuse scattering signal [32]. Spin correlations can then be obtained from the isolated magnetic diffuse scattering by using Reverse Monte Carlo (RMC) methods [33,34] or by least-squares fitting to analytical formulae [12,35]. The RMC methods are highly effective, but they do not provide information on the underlying magnetic interactions driving the spin correlations. The gold standard in the field for determining the interactions is inelastic neutron scattering, where the spin-wave spectra in the magnetically ordered state is measured and then fitted using linear spin-wave theory [36]. However, the exchange interactions can also be obtained from magnetic diffuse scattering in the paramagnetic state by using mean-field Onsager reaction-field theory [37–40]. This approach allows for the least-squares fitting of the exchange constants of a desired magnetic Hamiltonian. The general purpose software SPINTERACT for Onsager reaction-field fitting has recently become available [41,42].

Here we report on the magnetic diffuse scattering of the $S$ = 5/2 fcc antiferromagnets Ba$_2$MnTeO$_6$ and Ba$_2$MnWO$_6$ above their respective magnetic ordering temperatures of 20 K and 8 K. We show that both materials have significant short-range spin correlations in the paramagnetic state up to at least $T$ = 7$T_N$. The nearest-neighbor spin correlations in Ba$_2$MnTeO$_6$ are antiferromagnetic and next-nearest-neighbor correlations are ferromagnetic. For Ba$_2$MnWO$_6$, we find both nearest-neighbor and next-nearest-neighbor spin correlations to be antiferromagnetic. We also show that Onsager reaction-field fitting of diffuse scattering in the paramagnetic state can produce comparable exchange constants to linear spin wave theory fits of inelastic neutron scattering in the ordered state. We obtain $J_1$ = -3.25(3) K and $J_2$ = 0.41(2) K for $Ba_2MnTeO_6$ and $J_1$ = -1.08(1) K and $J_2$ = -0.88(1) K for $Ba_2MnWO_6$.

## II. EXPERIMENTAL

Our experiments were performed on polycrystalline powder samples of $Ba_2MnTeO_6$ and $Ba_2MnWO_6$ that were previously synthesized and characterized in refs. [17,18]. $Ba_2MnTeO_6$ and $Ba_2MnWO_6$ were prepared by conventional solid-state reaction methods. Stoichiometric quantities of $BaCO_3$ (99.997%), $MnO_2$ (99.999%), $TeO_2$ (99.9995%) and $WO_3$ (99.998%) were ground in an agate mortar. $Ba_2MnTeO_6$ was first calcined in air at 900 °C with the synthesis carried out in air at 1100 °C for 96h. $Ba_2MnWO_6$ was calcined in air at 800 °C and the synthesis was carried out in 5% $H_2/N_2$ at 1250 °C for 96h. These samples were characterised by neutron diffraction in refs. [17,18] and found to be of high quality. Our $Ba_2MnWO_6$ sample is phase pure based on neutron and laboratory X-ray measurements [18]. Our $Ba_2MnTeO_6$ sample [17] has a 1.0(1)% 2H-$BaMnO_3$ impurity [45] and a trace $Mn_3O_4$ impurity detectable only by magnetometry. The crystal and magnetic structures were visualized using VESTA [46].

Magnetic susceptibilities as function of temperature were measured on a Quantum Design MPMS3 magnetometer. Approximately 100 mg of sample powder was enclosed in a gelatin capsule and placed in a plastic straw sample holder. The measurements were carried out in DC mode under an applied field of 1000 Oe in the temperature range 2-300 K. The sample shape was estimated as a cylinder with diameter of 5mm and a height of 2mm leading to a sample moment artefact factor of 1.072 for DC measurements. The measured magnetization was divided by this value.

Polarized neutron scattering experiments were performed at the D7 diffuse scattering spectrometer at the Institut Laue-Langevin [47,48]. Approximately 6–8 g of sample powder was packed in aluminum cans with inserts forming an annulus shape. The incoming neutron energy was 3.55 meV corresponding to a wavelength of 4.8 Å. $Ba_2MnTeO_6$ with $T_N$ = 20 K was measured at 30 K, 45 K, 60 K, 100 K and 150 K. $Ba_2MnWO_6$ with $T_N$ = 8 K was measured at 13 K, 18 K, 30 K, 40 K, 60 K and 100 K. Data reduction was carried out using LAMP [49]. The data were corrected for detector efficiency using a vanadium standard and for polarization quality using a quartz standard. To estimate the background, an empty can and cadmium were measured as well.

The magnetic diffuse scattering was extracted using xyz polarization analysis [50]. The measured diffuse scattering was normalized to absolute units (barns sr$^{-1}$ f.u.$^{-1}$) initially by using a vanadium standard and finally by using the nuclear Bragg scattering as an internal standard. The nuclear Bragg scattering at 100 K was refined using FULLPROF [51]. The magnetic diffuse scattering was normalized by dividing by:

$$s = S_{\text{FP}} \times \frac{2\pi^2 ZV}{45\lambda^3}$$

where $S_{\text{FP}}$ is the scale factor obtained from the nuclear structure refinement on FULLPROF, $V$ is the unit cell volume, $Z$ is the number of formula units per unit cell and $\lambda$ is the neutron wavelength.

The magnetic diffuse scattering was modelled using a Reverse Monte Carlo (RMC) method as implemented in SPINVERT [33,34]. This approach is entirely independent of the underlying magnetic Hamiltonian. We used 8 × 8 × 8 supercells with 2048 spins for both compounds. The $Mn^{2+}$ spins were modelled as Heisenberg spins without restrictions on their direction. The analysis was repeated 10 times for each dataset to reduce statistical noise. The radial spin correlation functions were obtained from the RMC spin configurations using SPINCORREL [34]. The spin correlation length $\xi$ was estimated by fitting the exponential decay of spin correlations for pairs of spins along the same direction.

The magnetic diffuse scattering was also modelled using mean-field Onsager reaction-field theory [37–40] with SPINTERACT [42]. This allowed us to extract the interaction parameters for a simple $J_1$-$J_2$ Heisenberg model. All magnetic diffuse scattering datasets and magnetic susceptibility were fitted simultaneously. The analysis was repeated multiple times starting from random values of $J_1$ and $J_2$ to avoid local minima.

Our experimental neutron scattering data are available online at ref. [52].

## III. RESULTS

### A. Magnetic susceptibility

Magnetic susceptibilities of $Ba_2MnTeO_6$ and $Ba_2MnWO_6$ are shown in Figure 2. Both materials are antiferromagnetic with the magnetic transitions occurring at $T_N$ = 20 K and $T_N$ = 8 K, respectively. Our $Ba_2MnTeO_6$ sample has a trace $Mn_3O_4$ impurity, which is too small to be detected by laboratory X-ray diffraction. This $Mn_3O_4$ impurity is responsible for the ferrimagnetic transition around 44 K and the minor divergence of the zero-field cool (ZFC) and field-cool (FC) curves for the $Ba_2MnTeO_6$ sample. For $Ba_2MnWO_6$, the ZFC and FC curves overlap completely.

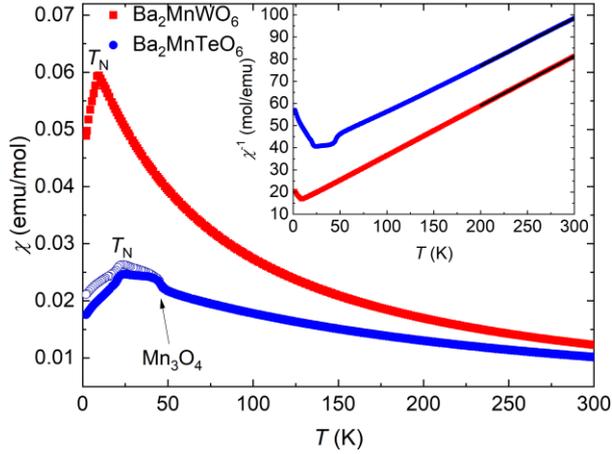

Figure 2. (a) Magnetic susceptibilities of $Ba_2MnTeO_6$ and $Ba_2MnWO_6$ measured with an applied field of 1000 Oe. The zero-field cooled curve is shown with filled symbols and the field-cool curve with empty symbols. A clear antiferromagnetic transition is observed in $Ba_2MnWO_6$ at $T_N$ = 8 K. An antiferromagnetic transition is observed at $T_N \approx 20$ K in $Ba_2MnTeO_6$ along with a trace ferrimagnetic $Mn_3O_4$ impurity, which causes a small divergence in the zero-field cool and field-cool curves. (b) Curie-Weiss fits to inverse magnetic susceptibility of $Ba_2MnTeO_6$ and $Ba_2MnWO_6$ in the range $T$ = 200-300 K. We obtain $\mu_{eff}$ = 6.09(1) $\mu_B$ and $\vartheta_{CW}$ = -156(1) K for $Ba_2MnTeO_6$ and $\mu_{eff}$ = 5.97(1) $\mu_B$ and $\vartheta_{CW}$ = -63(1) K for $Ba_2MnWO_6$, respectively.

The inverse magnetic susceptibilities were fitted to the Curie-Weiss law between 200 K and 300 K. For $Ba_2MnTeO_6$, we obtain an effective paramagnetic moment of $\mu_{eff}$ = 6.09(1) $\mu_B$ and a Curie-Weiss constant of $\vartheta_{CW}$ = -156(1) K. Similarly, for $Ba_2MnWO_6$ the fitting yields $\mu_{eff}$ = 5.97(1) $\mu_B$ and $\vartheta_{CW}$ = -63(1) K. The effective paramagnetic moments are slightly larger than the expected spin-only value of $\mu_{SO}$ = 5.92 $\mu_B$. The negative Curie-Weiss temperatures are consistent with both materials being antiferromagnetic, and the overall antiferromagnetic interactions are stronger in $Ba_2MnTeO_6$ than in $Ba_2MnWO_6$.

We have previously reported the magnetic susceptibility of these samples measured in SQUID-VSM mode on the MPMS3 [17,18], but we did not consider sample shape effects at the time. These are more prominent in SQUID-VSM mode than in the traditional DC mode, and lead to an overestimation of the magnetization. As a result, the effective paramagnetic moments were overestimated in our previous measurements with $\mu_{eff} \approx 6.3$ $\mu_B$. The fitted Curie-Weiss constants are unaffected by this issue and are identical to our previous results. The effective paramagnetic moments and Curie-Weiss constants are consistent with other reports on these materials [20,26,28].

## B. Magnetic diffuse scattering and spin correlations

The magnetic diffuse scattering of Ba$_2$MnTeO$_6$ and Ba$_2$MnWO$_6$ was investigated in the paramagnetic state above their $T_N$ = 20 K and 8 K, respectively. The magnetic diffuse scattering of Ba$_2$MnTeO$_6$ at selected temperatures is shown in Figure 3(a). The diffuse scattering at 30 K has two broad peaks at $|Q|$ = 0.78 Å$^{-1}$ and $|Q|$ = 1.8 Å$^{-1}$. The main peak at $|Q|$ = 0.78 Å$^{-1}$ arises from the (001) reflection of the ordered Type I antiferromagnetic structure, whereas the peak at $|Q|$ = 1.8 Å$^{-1}$ is related to the magnetic reflections (201) and (112). This confirms that the short-range correlated state above $T_N$ is related to the magnetic order below $T_N$. As expected, these features become weaker and broader as temperature is increased to 60 K and 100 K. At 100 K the second peak at $|Q|$ = 1.8 Å$^{-1}$ can no longer be observed. It should be noted that the main (001) peak is present in the 150 K dataset, which reveals that short-range spin correlations persist at least up to this temperature. In the absence of spin correlations, the scattering would simply follow the magnetic form factor of Mn$^{2+}$.

The magnetic diffuse scattering of Ba$_2$MnWO$_6$ at selected temperatures is shown in Figure 3(b). The magnetic diffuse scattering at 13 K has a main peak at $|Q|$ = 0.68 Å$^{-1}$, which arises from the ($\frac{1}{2}\frac{1}{2}\frac{1}{2}$) reflection of the Type II antiferromagnetic order below $T_N$. This peak becomes broader and weaker as temperature is increased to 30 K and 60 K as expected. A second weak peak at $|Q| \approx$ 1.95 Å$^{-1}$ related to the ($\frac{3}{2}\frac{3}{2}\frac{3}{2}$) reflection is observed in our 13 K and 18 K data, but not at 30 K or higher temperatures. The main peak is observed up to at least 100 K, which confirms that spin correlations are present far above $T_N$ = 8 K for Ba$_2$MnWO$_6$. Plots of the magnetic diffuse scattering for both compounds at all temperatures measured are provided in the Supplemental Material [53].

The magnetic diffuse scattering of Ba$_2$MnTeO$_6$ and Ba$_2$MnWO$_6$ was analysed with an RMC approach using SPINVERT. In this method, configurations of spins in a large supercell are randomly adjusted to improve agreement with the measured magnetic diffuse scattering. The magnetic diffuse scattering from the spin configurations can be calculated from [12]:

$$\left(\frac{d\sigma}{d\Omega}\right)_{mag}(Q) = (\gamma_n r_0)^2 \left(\frac{1}{2}F(Q)\right)^2 \mu_{\text{eff}}^2 \left\{\frac{2}{3} + \frac{1}{N}\sum_{i,j}\left[A_{ij}\frac{\sin Qr_{ij}}{Qr_{ij}} + B_{ij}\left(\frac{\sin Qr_{ij}}{(Qr_{ij})^3} - \frac{\cos Qr_{ij}}{(Qr_{ij})^3}\right)\right]\right\} \quad (2)$$

where $A_{ij} = S_i^x S_j^y$, $B_{ij} = 2S_i^z S_j^z - S_i^x S_j^x$, $F(Q)$ is the magnetic form factor of Mn$^{2+}$, $\mu_{\text{eff}}^2 = g^2 S(S+1)$ and the sums are taken over all spin pairs at radial distance $r_{ij}$ up to the maximum distance of half of the side of the supercell (32 Å).

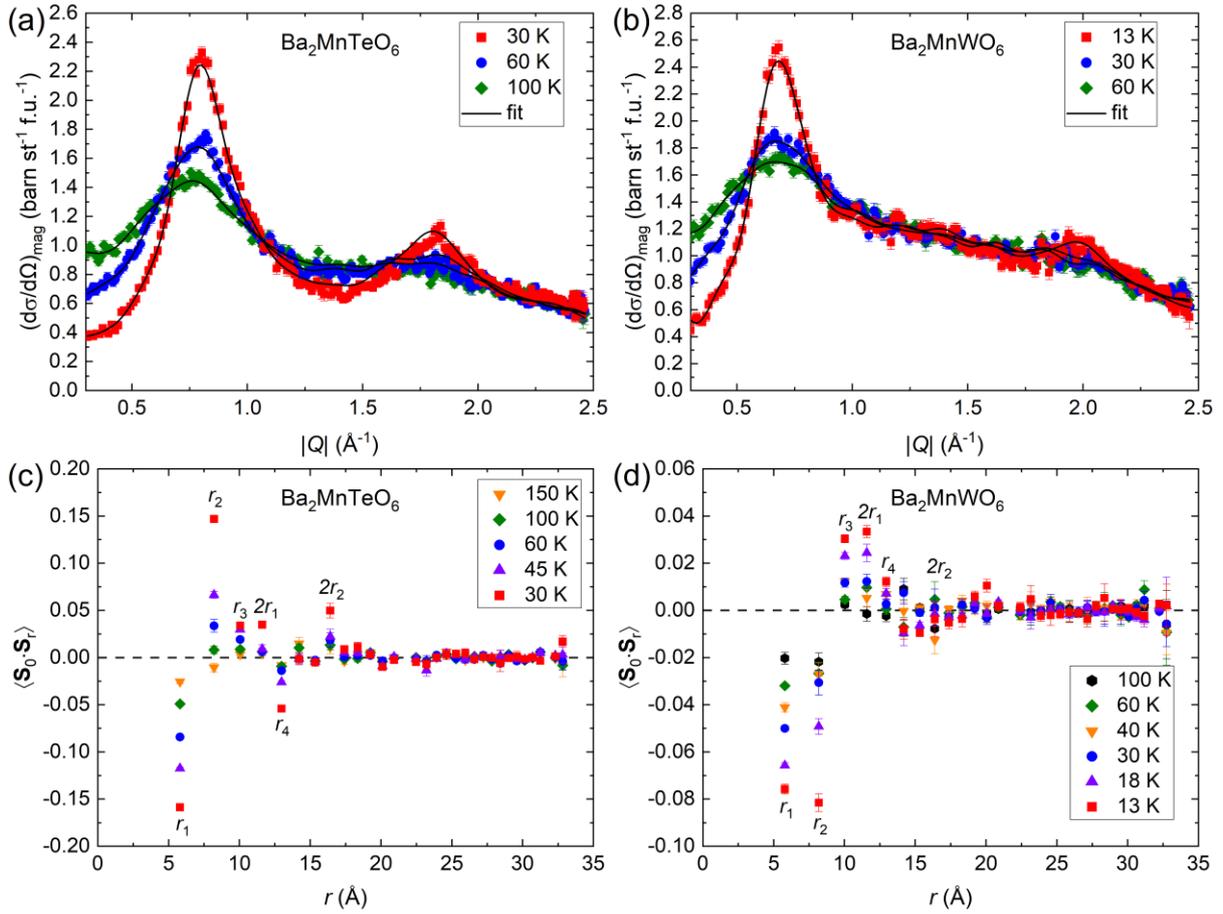

Figure 3. Magnetic diffuse scattering of (a) $Ba_2MnTeO_6$ and (b) $Ba_2MnWO_6$ above $T_N$ fitted using an RMC method. The maximum occurs at $|Q| \approx 0.78$ Å$^{-1}$ for $Ba_2MnTeO_6$ and $|Q| \approx 0.68$ Å$^{-1}$ for $Ba_2MnWO_6$. The radial spin correlation functions obtained from the RMC fits for (c) $Ba_2MnTeO_6$ and (d) $Ba_2MnWO_6$. The spin correlations for $Ba_2MnTeO_6$ are as expected for Type I order with an antiferromagnetic $\langle S_0 \cdot S_1 \rangle$ and ferromagnetic $\langle S_0 \cdot S_2 \rangle$. Both $\langle S_0 \cdot S_1 \rangle$ and $\langle S_0 \cdot S_2 \rangle$ are antiferromagnetic for $Ba_2MnWO_6$, while one would expect $\langle S_0 \cdot S_1 \rangle = 0$ due to the Type II ordering. The antiferromagnetic $\langle S_0 \cdot S_1 \rangle$ is driven by the substantial antiferromagnetic $J_1$ interaction in $Ba_2MnWO_6$.

The resulting fits for $Ba_2MnTeO_6$ and $Ba_2MnWO_6$ are shown as the black lines in Figure 3(a) and Figure 3(b). Our SPINVERT analysis reproduces the main features of magnetic diffuse scattering for both compounds at all temperatures measured. The main peaks at $|Q| = 0.78$ Å$^{-1}$ for $Ba_2MnTeO_6$ and $|Q| = 0.68$ Å$^{-1}$ for $Ba_2MnWO_6$ are very well described by the SPINVERT fits. For $Ba_2MnWO_6$, our model includes a number of smaller peaks between $|Q| = 1 - 2$ Å$^{-1}$ that might be signs of overfitting. Radial spin correlation functions for both compounds were calculated from the RMC-modelled spin configurations. The spin correlations were normalised such that $\langle S_0 \cdot S_r \rangle = 1$ corresponds to complete ferromagnetic alignment with neighboring spins at distance $r$, and conversely -1 corresponds to complete antiferromagnetic alignment. The expected nearest-neighbor and next-nearest-neighbor spin correlations for Type I antiferromagnetic order are $\langle \mathbf{S}_0 \cdot \mathbf{S}_1 \rangle = -1/3$ and $\langle \mathbf{S}_0 \cdot \mathbf{S}_2 \rangle = 1$ and for Type II order $\langle \mathbf{S}_0 \cdot \mathbf{S}_1 \rangle = 0$ and $\langle \mathbf{S}_0 \cdot \mathbf{S}_2 \rangle = -1$.

The radial spin correlation functions for $Ba_2MnTeO_6$ and $Ba_2MnWO_6$ at various temperatures are shown in Figure 3(c) and Figure 3(d), respectively. For both compounds, the nearest-neighbor and next-nearest-neighbor correlations are distinctly stronger than further-neighbor correlations. As expected, we observe an antiferromagnetic $\langle \mathbf{S}_0 \cdot \mathbf{S}_1 \rangle$ and a ferromagnetic $\langle \mathbf{S}_0 \cdot \mathbf{S}_2 \rangle$ for $Ba_2MnTeO_6$. We estimate the spin correlation length at 30 K to be $\xi$ = 3.9(1) Å. The spin correlations become weaker with increasing temperature without significant shifts in relative magnitudes, with the exception of a weak ferromagnetic $\langle \mathbf{S}_0 \cdot \mathbf{S}_2 \rangle$ in the 150 K dataset. This shift from ferromagnetic to weakly antiferromagnetic $\langle \mathbf{S}_0 \cdot \mathbf{S}_2 \rangle$ at high temperatures was also observed in the isostructural Type I antiferromagnet $Ba_2YRuO_6$ [44].

For $Ba_2MnWO_6$, $\langle \mathbf{S}_0 \cdot \mathbf{S}_1 \rangle$ and $\langle \mathbf{S}_0 \cdot \mathbf{S}_2 \rangle$ are both antiferromagnetic, whereas $\langle \mathbf{S}_0 \cdot \mathbf{S}_1 \rangle = 0$ is expected from the Type II magnetic structure. This is not a discrepancy, but related to the underlying magnetic interactions of $Ba_2MnWO_6$. Despite the Type II structure, the nearest-neighbor interaction $J_1$ is actually stronger than $J_2$ in this material [18]. This is reflected in the significant antiferromagnetic $\langle \mathbf{S}_0 \cdot \mathbf{S}_1 \rangle$ correlations observed in the magnetic diffuse scattering above $T_N$. Overall, the spin correlations are weaker in $Ba_2MnWO_6$ than in $Ba_2MnTeO_6$, which is consistent with the stronger antiferromagnetic interactions in $Ba_2MnTeO_6$ [17,18]. The spin correlation length at 13 K was estimated to be $\xi$ = 2.8(6) Å. The spin correlations of $Ba_2MnWO_6$ become weaker with increasing temperature as expected, but there is also a change in the relative strengths of the antiferromagnetic $\langle \mathbf{S}_0 \cdot \mathbf{S}_1 \rangle$ and $\langle \mathbf{S}_0 \cdot \mathbf{S}_2 \rangle$ correlations. At 13 K, just above $T_N$ = 8 K, $\langle \mathbf{S}_0 \cdot \mathbf{S}_1 \rangle$ and $\langle \mathbf{S}_0 \cdot \mathbf{S}_2 \rangle$ are equally strong within error. Between 18 K and 60 K, the nearest-neigbhor $\langle \mathbf{S}_0 \cdot \mathbf{S}_1 \rangle$ is stronger than $\langle \mathbf{S}_0 \cdot \mathbf{S}_2 \rangle$, whereas at 100 K both correlations are within error of each other again.

The spin correlations from magnetic diffuse scattering are typically obtained in the quasi-static approximation. For this to hold, energy integration should be carried out over the full inelastic spectrum [32]. Our relatively low incoming neutron energy of 3.55 meV poses a problem for the energy integration of $Ba_2MnTeO_6$, which has stronger magnetic interactions than $Ba_2MnWO_6$ and inelastic scattering up to higher energies [17,18]. We can estimate this effect by evaluating the effective paramagnetic moments obtained from fitting the magnetic diffuse scattering, where $\mu_{eff}^2$ is used as a scale factor in equation (2). We obtain $\mu_{eff}$ = 5.16 $\mu_B$ for $Ba_2MnTeO_6$ and $\mu_{eff}$ = 5.73 $\mu_B$ for $Ba_2MnWO_6$. These are slightly smaller than obtained from the Curie-Weiss fits or the spin-only value of $\mu_{SO}$ = 5.92 $\mu_B$. This suggests we capture approximately 76% of the full spectral weight for $Ba_2MnTeO_6$ and 93% for $Ba_2MnWO_6$.

## C. Magnetic interactions from Onsager reaction-field theory

Single crystal magnetic diffuse scattering for Heisenberg spins within Onsager reaction-field theory is given by [42]:

$$\left(\frac{d\sigma}{d\Omega}\right)_{mag}(\mathbf{Q}) = \frac{2}{3}(\gamma_n r_0)^2 \left(\frac{1}{2}F(Q)\right)^2 \mu_{\text{eff}}^2 \times \frac{1}{N}\sum_{\mu=1}^{N}\frac{\left|\sum_i U_{i\mu}(\mathbf{Q})e^{i\mathbf{Q}\cdot\mathbf{R}_i}\right|^2}{1-\chi_0[\lambda_\mu(\mathbf{Q})-\lambda]} \quad (3)$$

where $\lambda_\mu(\mathbf{Q})$ are the eigenvalues of the interaction matrix of the spin Hamiltonian, $U_{i\mu}(\mathbf{Q})$ are its eigenvectors components and $\lambda$ is the reaction field, which describes the deviation from the mean-field due to local spin correlations. The magnetic ordering temperature $T_N$ can be estimated as the highest temperature, where the denominator $1-\chi_0[\lambda_\mu(\mathbf{Q})-\lambda]$ is zero for any wavevector and mode [42]. Similarly, the propagation vector of the magnetically ordered phase below $T_N$ is given by the $\mathbf{Q}$ corresponding to the maximum eigenvalue $\lambda_{\max}(\mathbf{Q})$ [42]. The single crystal diffuse scattering from equation (3) was powder-averaged to fit our powder magnetic diffuse scattering data. Magnetic susceptibility measurements probe the $\mathbf{Q}$ = 0 wavevector, and the isotropic (powder) susceptibility can be calculated from [39,42]:

$$\chi T = \frac{\mu_{\text{eff}}^2}{N}\sum_{\mu=1}^{N}\frac{\left|\sum_i U_{i\mu}(\mathbf{0})\right|^2}{1-\chi_0[\lambda_\mu(\mathbf{0})-\lambda]} \quad (4)$$

Magnetic interactions can be fitted using a non-linear least-squares method by first calculating the interaction matrix from the exchange constants, then calculating the reaction field at each temperature and finally calculating the magnetic diffuse scattering for each dataset from equation (3) and the magnetic susceptibility from equation (4). All datasets were fitted simultaneously to obtain the final exchange constants.

The spin wave spectra of $Ba_2MnTeO_6$ and $Ba_2MnWO_6$ are well described by a simple fcc $J_1$-$J_2$ Heisenberg model [17,18]. As such, the $J_1$-$J_2$ model provided a natural starting point for our fitting. Onsager reaction-field fits of the magnetic diffuse scattering and magnetic susceptibility for $Ba_2MnTeO_6$ are shown in Figure 4. The fitting captures the main features of the data for all datasets, but does not provide as good a fit as the RMC approach due to the highly constrained nature of the fitting. Notably, the model underestimates the intensity of the main peak at $|Q|$ = 0.78 Å$^{-1}$ for $T$ = 30 K, 45 K and 60 K and underestimates the intensity in the low-$Q$ region $|Q|$ = 0.3-0.6 Å$^{-1}$ at 100 K and 150 K. The fitting assumes that the experiments capture the full neutron spectral weight, and these minor issues could be related to breakdown of the quasi-static approximation in our magnetic diffuse scattering data for $Ba_2MnTeO_6$.

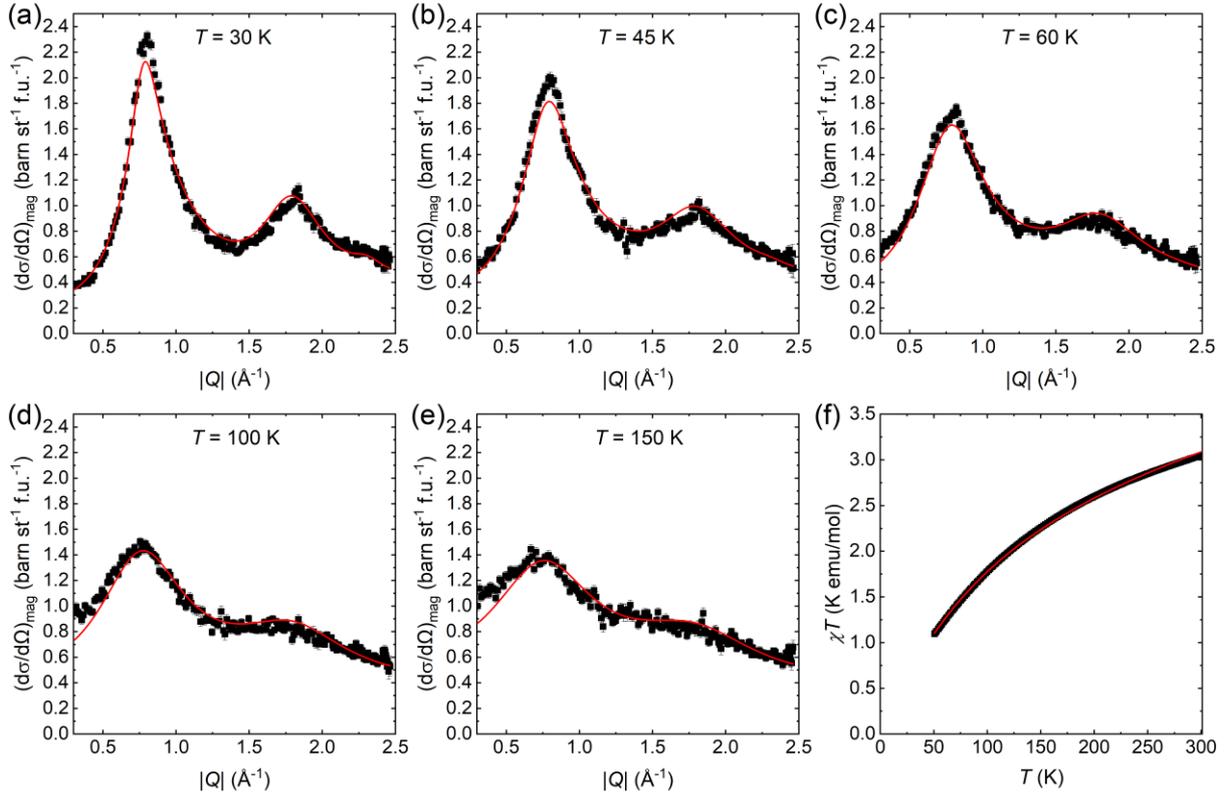

Figure 4. Magnetic diffuse scattering of Ba$_2$MnTeO$_6$ above $T_N$ (black squares) and fits to a $J_1$-$J_2$ Heisenberg model using mean-field Onsager reaction-field theory (red lines). Data collected at temperatures (a) 30 K, (b) 45 K, (c) 60 K, (d) 100 K and (e) 150 K. (f) Magnetic susceptibility (ZFC) plotted as $\chi T$ vs $T$. All datasets were fitted simultaneously to obtain $J_1$ = -3.25(3) K and $J_2$ = 0.41(2) K.

The obtained exchange interactions from Onsager reaction-field fits for Ba$_2$MnTeO$_6$ are $J_1$ = -3.25(3) K and $J_2$ = 0.41(2) K. These values are very close to the $J_1$ = -3.95 K and $J_2$ = 0.35 K from linear spin wave theory modeling of the inelastic neutron scattering spectra [17]. The predicted ordering temperature $T_N$ = 17.4 K is close to the experimental value of 20 K [17]. The fitting also correctly predicts the magnetic propagation vector **k** = (0, 0, 1) [17] corresponding to Type I order from the maximum eigenvalue $\lambda_{max}(\mathbf{Q})$. Moreover, the extracted exchange interactions also place Ba$_2$MnTeO$_6$ in the Type I region of the fcc $J_1$-$J_2$ phase diagram with an antiferromagnetic $J_1$ and a ferromagnetic $J_2$ [9,54]. We also tested fitting each temperature separately to further investigate possible energy integration issues (Supplemental Material [53]). The obtained exchange interactions are closest to the linear spin wave theory results [17] at low temperatures, and then start to systematically deviate with increasing temperature. This is consistent with energy integration issues, which are most noticeable at high temperatures as the linewidth becomes broader and intensity shifts towards low-$|Q|$ [32].

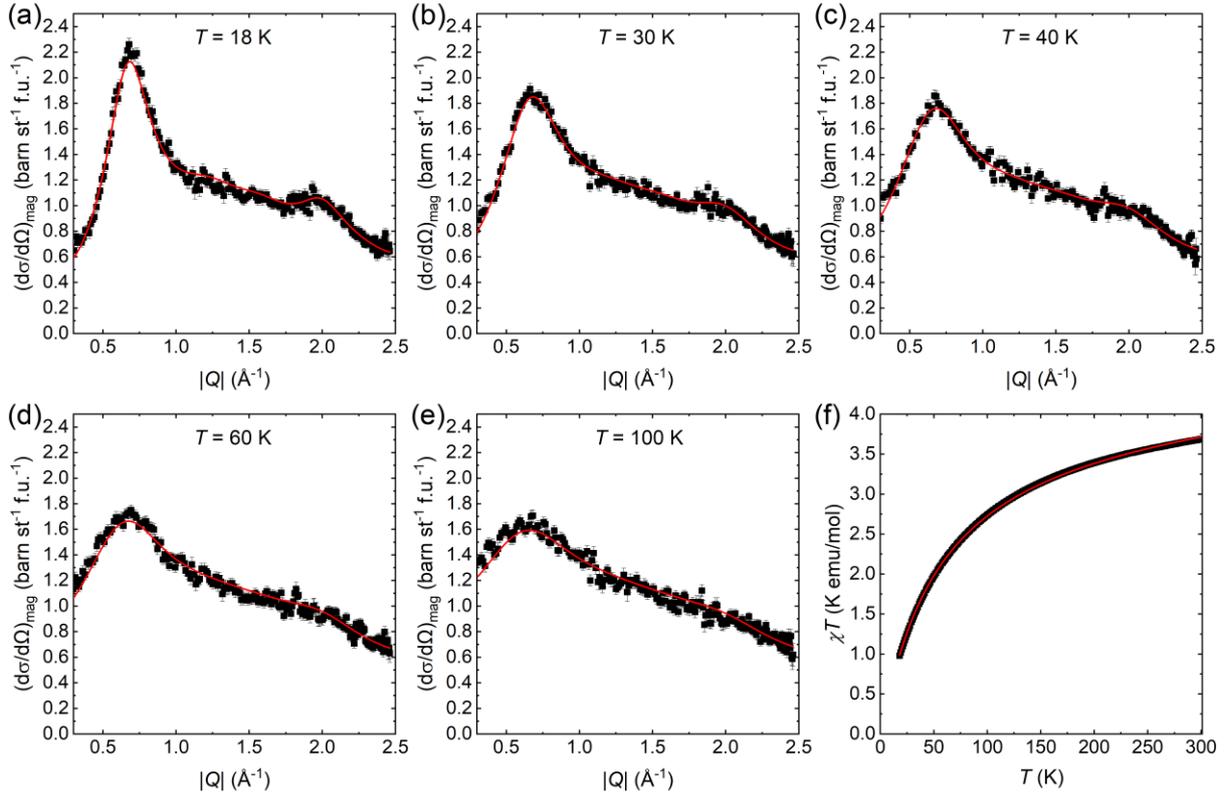

Figure 5. Magnetic diffuse scattering of $Ba_2MnWO_6$ above $T_N$ (black squares) and fits to a $J_1$-$J_2$ Heisenberg model using mean-field Onsager reaction-field theory (red lines). Data collected at temperatures (a) 18 K, (b) 30 K, (c) 40 K, (d) 60 K and (e) 100 K. (f) Magnetic susceptibility (ZFC) plotted as $\chi T$. All datasets were fitted simultaneously to obtain $J_1$ = -1.08(1) K and $J_2$ = -0.88(1) K.

Onsager reaction-field fits for $Ba_2MnWO_6$ are shown in Figure 5. The fitting describes the magnetic diffuse scattering of $Ba_2MnWO_6$ well at all temperatures, although the main peak at $|Q|$ = 0.68 Å$^{-1}$ is missing some intensity in the 18 K fit. It should be noted that Onsager reaction-field theory is an extension of mean-field theory, and as such it is accurate at high temperatures. Therefore, minor issues in the low-$T$ datasets are to be expected. For the same reason, we did not include the lowest temperature dataset at 13 K in our fitting. We do not observe the low-$|Q|$ intensity issues in the $Ba_2MnWO_6$ fits at high temperatures that were found in the $Ba_2MnTeO_6$ fitting. This is to be expected, as we capture almost the full spectral weight for $Ba_2MnWO_6$.

The exchange interactions for $Ba_2MnWO_6$ from the fitting are $J_1$ = -1.08(1) K and $J_2$ = -0.88(1) K. Our Onsager reaction-field results are almost identical to linear spin wave theory, which yields $J_1$ = -0.93 K and $J_2$ = -0.88 K [18]. Moreover, the fitting gives a very good estimate of the magnetic ordering temperature as $T_N$ = 7.7 K, which is consistent with the $T_N$ = 8 K observed experimentally. The propagation vector also is correctly predicted [18] as **k** = (1/2, 1/2, 1/2) from the maximum eigenvalue corresponding to Type II order. This is also consistent with the obtained exchange interactions and the phase diagram of the $J_1$-$J_2$ fcc model: $J_1$ and $J_2$ are both antiferromagnetic and $J_2/J_1 > 0.5$ fulfilling the criteria for Type II order [9,54].

We also tested whether further-neighbor magnetic interactions are relevant for $Ba_2MnTeO_6$ and $Ba_2MnWO_6$. Our results are summarized in Table I. Including a weak ferromagnetic third-neighbor $J_3$ interaction leads to a slight improvement in the fit for $Ba_2MnTeO_6$ with $J_3$ refining to 17% of the value of $J_2$. This suggest $J_3$ could be relevant for $Ba_2MnTeO_6$, although $J_1$ and $J_2$ do not significantly change from their values for the $J_1$-$J_2$ fit. For $Ba_2MnWO_6$, including a $J_3$ does not lead to any improvement in the fit and $J_3$ refines to almost nothing. This suggests further-neighbor interactions are very weak for $Ba_2MnWO_6$.

Table I. Exchange interactions obtained from Onsager reaction-field theory fitting of magnetic diffuse scattering. The addition of a third-neighbor Heisenberg interaction $J_3$ slightly improves the fit for $Ba_2MnTeO_6$, but not for $Ba_2MnWO_6$.

|  | $J_1$ (K) | $J_2$ (K) | $J_3$ (K) | $R_{wp}$ (%) |
|---|---|---|---|---|
| $Ba_2MnTeO_6$ $J_1$-$J_2$ | -3.25(3) | 0.41(2) | - | 6.34 |
| $Ba_2MnTeO_6$ $J_1$-$J_2$-$J_3$ | -2.93(4) | 0.69(3) | 0.12(1) | 6.21 |
| $Ba_2MnWO_6$ $J_1$-$J_2$ | -1.08(1) | -0.88(1) | - | 4.26 |
| $Ba_2MnWO_6$ $J_1$-$J_2$-$J_3$ | -1.05(2) | -0.88(2) | 0.02(1) | 4.25 |

**IV. DISCUSSION**

Previous inelastic neutron scattering and muon spin rotation and relaxation experiments on $Ba_2MnTeO_6$ and $Ba_2MnWO_6$ have revealed the presence of short-range correlated magnetism above $T_N$ in both materials [17,18,28]. Here, we have characterised the spin correlations in these states using RMC analysis of magnetic diffuse scattering data collected above $T_N$. We find strong antiferromagnetic nearest-neighbor correlations $\langle \bm{S}_0 \cdot \bm{S}_1 \rangle$ and ferromagnetic next-nearest-neighbor correlations $\langle \bm{S}_0 \cdot \bm{S}_2 \rangle$ for $Ba_2MnTeO_6$ at $T$ = 30 – 150 K. This confirms that short-range Type I correlations persist even in the paramagnetic state. The spin correlations become weaker with increasing temperature, but persist up to at least 150 K. For $Ba_2MnWO_6$, we find antiferromagnetic $\langle \bm{S}_0 \cdot \bm{S}_1 \rangle$ and $\langle \bm{S}_0 \cdot \bm{S}_2 \rangle$ correlations at $T$ = 13 – 100 K. These are not strictly Type II correlations, where $\langle \bm{S}_0 \cdot \bm{S}_1 \rangle = 0$ would be expected. Instead, the antiferromagnetic $\langle \bm{S}_0 \cdot \bm{S}_1 \rangle$ correlations arise from the strong antiferromagnetic $J_1$ interaction in $Ba_2MnWO_6$ [18]. This shows that the magnetic frustration related to the nearest-neighbor interactions in the ordered state is lifted in the paramagnetic state [13]. The short-range spin correlations persist up to 100 K.

The natural comparison for the Mn$^{2+}$ double perovskites investigated here is the archetypical $S$ = 5/2 fcc antiferromagnet MnO. Paddison *et al.* [13] investigated the magnetic diffuse scattering of a MnO single crystal in the paramagnetic state at 160 K using RMC methods. The spin correlations are dominated by antiferromagnetic next-nearest-neighbor correlations $\langle \boldsymbol{S}_0 \cdot \boldsymbol{S}_2 \rangle$, while the nearest-neighbor $\langle \boldsymbol{S}_0 \cdot \boldsymbol{S}_1 \rangle$ is also antiferromagnetic. The spin correlations above $T_N$ in MnO are very similar to our results for Ba$_2$MnWO$_6$, although we find stronger antiferromagnetic $\langle \boldsymbol{S}_0 \cdot \boldsymbol{S}_1 \rangle$ likely arising from the relatively stronger $J_1$ interaction in Ba$_2$MnWO$_6$ [18]. Further analysis of the single crystal diffuse scattering of MnO revealed local order in small domains each associated with one of four symmetry-equivalent (1/2, 1/2, 1/2) periodicities along body diagonals [13]. We are unable to evaluate the domain structure in Ba$_2$MnWO$_6$ or Ba$_2$MnTeO$_6$, as we are limited to powder samples. Short-range correlated magnetism above $T_N$ has also been observed in the related fcc antiferromagnets CoO and NiO [55,56].

A few examples of short-range correlated magnetism in double perovskites are also known. Ba$_2$YRuO$_6$ is a cubic Ru$^{5+}$ double perovskite with an fcc lattice of $S$ = 3/2 spins [57], where two magnetic transitions are observed at $T_{N1}$ = 47 K and $T_{N2}$ = 36, respectively [44]. A Type I antiferromagnetic structure with an unusual noncoplanar 3-**q** spin texture develops below $T_{N2}$ = 36 K [58]. The magnetic diffuse scattering above $T_{N1}$ is similar to our observations for Ba$_2$MnTeO$_6$ with a broad peak at $|Q|$ = 0.75 Å arising from the (001) reflection of the Type I structure [44]. This diffuse scattering due to short-range magnetism is observed up to at least 200 K. At 50 K and 70 K, the spin correlations are consistent with the Type I structure with antiferromagnetic $\langle \boldsymbol{S}_0 \cdot \boldsymbol{S}_1 \rangle$ and ferromagnetic $\langle \boldsymbol{S}_0 \cdot \boldsymbol{S}_2 \rangle$. However, the next-nearest neighbor $\langle \boldsymbol{S}_0 \cdot \boldsymbol{S}_2 \rangle$ becomes antiferromagnetic at 100 K ($T \approx 2T_N$). While we observed a similar change in $\langle \boldsymbol{S}_0 \cdot \boldsymbol{S}_2 \rangle$ for Ba$_2$MnTeO$_6$, it only occurred at 150 K ($T \approx 7.5T_N$) – a significantly higher temperature when taking into account the energy scale of the magnetic interactions.

We have recently shown that the tetragonal Cu$^{2+}$ double perovskites Sr$_2$CuTeO$_6$ and Sr$_2$CuWO$_6$ display short-range correlated magnetism above $T_N$ [43]. Magnetism in these compounds is highly two-dimensional due to a Jahn-Teller distortion on the Cu$^{2+}$ and the accompanying orbital ordering [59]. Consequently, these materials are best described as $J_1$-$J_2$ square-lattice antiferromagnets [60,61]. Sr$_2$CuTeO$_6$ orders with a Néel antiferromagnetic structure below $T_N$ = 29 K [62], while Sr$_2$CuWO$_6$ orders with a Type II structure similar to Ba$_2$MnWO$_6$ below $T_N$ = 24 K [63,64]. Magnetic diffuse scattering at 40 K reveals significant in-plane spin correlations related to the magnetic order below $T_N$ for both compounds [43]. For Sr$_2$CuWO$_6$, $\langle \boldsymbol{S}_0 \cdot \boldsymbol{S}_1 \rangle \approx 0$ and the in-plane $\langle \boldsymbol{S}_0 \cdot \boldsymbol{S}_2 \rangle$ is antiferromagnetic as expected for Type II order. In contract, we observed antiferromagnetic $\langle \boldsymbol{S}_0 \cdot \boldsymbol{S}_1 \rangle$ for Ba$_2$MnWO$_6$. This difference is explained by the relative strengths of the exchange interactions: in $Sr_2CuWO_6$ $J_2$ is much stronger than $J_1$ with $J_2/J_1 \approx 8$, whereas in $Ba_2MnWO_6$ $J_1$ is actually slightly stronger than $J_2$ with $J_2/J_1 \approx 0.9$ [18,61].

$Ba_2MnTeO_6$ and $Ba_2MnWO_6$ are excellent test cases for Onsager reaction-field fitting of magnetic diffuse scattering data. In particular, the magnetic diffuse scattering in these materials is strong and easy to measure thanks to the high-spin $S = 5/2$ $Mn^{2+}$ magnetic cations. Moreover, these materials have a high-symmetry cubic structure with structurally well-ordered magnetic cations on a single crystallographic site. Finally, the magnetism is described by a simple fcc $J_1$-$J_2$ Hamiltonian, which is already known from inelastic neutron scattering studies [17,18]. In fact, we are able to directly compare exchange constants obtained by Onsager reaction-field fitting and linear spin wave theory using the same $Ba_2MnTeO_6$ and $Ba_2MnWO_6$ samples.

Our results on $Ba_2MnTeO_6$ and $Ba_2MnWO_6$ confirm that Onsager reaction-field fitting of magnetic diffuse scattering in the paramagnetic state can yield comparable exchange constants to linear spin-wave theory analysis of the inelastic neutron scattering in the ordered state. For $Ba_2MnTeO_6$ we obtained $J_1$ = -3.25(3) K and $J_2$ = 0.41(2) K from Onsager reaction-field fitting of magnetic diffuse scattering at T = 30 – 150 K. This is very close to our previous linear-spin wave theory analysis [17] of inelastic neutron scattering at $T$ = 7 K on the same sample, which yielded $J_1$ = -3.95 K and $J_2$ = 0.35 K. Additionally, Li *et al*. [26] have reported similar values of $J_1$ = -3.1(1) K and $J_2$ = 0.6(1) K based on inelastic neutron scattering measurements. Our Onsager fitting also raised the possibility of a weak third-neighbor interaction $J_3$ being relevant for $Ba_2MnTeO_6$. For $Ba_2MnWO_6$ we obtained $J_1$ = -1.08(1) K and $J_2$ = -0.88(1) K from Onsager fits at $T$ = 18 – 100 K. We have previously [18] measured the inelastic neutron scattering of this sample at $T$ = 2 K yielding nearly identical exchange constants of $J_1$ = -0.93 K and $J_2$ = -0.88 K.

Onsager reaction-field fitting has not been used widely in the analysis of magnetic diffuse scattering data likely due to the lack of general purpose fitting software until the recent release of SPINTERACT [42]. This approach has been shown to describe well the magnetic diffuse scattering of MnO between $T$ = 130 K and 220 K yielding reasonable exchange constants [65]. A direct comparison to linear spin wave theory is complicated by the structural distortion at $T_N$, which means that the crystal structure and magnetic interactions of MnO are slightly different below and above $T_N$. This is not an issue with $Ba_2MnTeO_6$ and $Ba_2MnWO_6$, which retain their cubic structure below $T_N$ [17,18,20]. Onsager reaction-field fitting has also been successfully applied to model frustrated system gadolinium gallium garnet $Gd_3Ga_5O_{12}$ [42], the pyrochlore antiferromagnets $LiGaCr_4S_8$ [66] and $Gd_2Pt2O_7$ [67], the candidate Kitaev material $NaNi_2BiO_{6-\delta}$ [41] and the Skyrmion material $Gd_2PdSi_3$ [68]. Our results add to the growing literature on Onsager reaction-field fitting of magnetic diffuse scattering and provide a direct comparison of Onsager fitting and linear spin wave theory for two Heisenberg antiferromagnets.

**V. CONCLUSIONS**

We investigated the magnetic diffuse scattering of the $S$ = 5/2 fcc antiferromagnets $Ba_2MnTeO_6$ and $Ba_2MnWO_6$ above their respective magnetic ordering transitions of 20 K and 8 K. Our results showed that short-range correlated magnetism occurs in both compounds above $T_N$ confirming previous reports [17,18,28]. Broad features related to the magnetic Bragg positions of the ordered phases were observed in the magnetic diffuse scattering. Reverse Monte Carlo analysis revealed antiferromagnetic nearest-neighbor $\langle \boldsymbol{S}_0 \cdot \boldsymbol{S}_1 \rangle$ and ferromagnetic next-nearest neighbor $\langle \boldsymbol{S}_0 \cdot \boldsymbol{S}_2 \rangle$ correlations for $Ba_2MnTeO_6$ at $T$ = 30 – 150 K. These correlations are related to the Type I order below $T_N$. We observe antiferromagnetic $\langle \boldsymbol{S}_0 \cdot \boldsymbol{S}_1 \rangle$ and $\langle \boldsymbol{S}_0 \cdot \boldsymbol{S}_2 \rangle$ correlations for $Ba_2MnWO_6$ at $T$ = 13 – 100 K. While related to the Type II order below $T_N$, the magnetic frustration of the antiferromagnetic nearest-neighbor interaction is partially lifted in the paramagnetic state. Short-range spin correlations were observed up to at least $T = 7T_N$ in both compounds.

The magnetic diffuse scattering was also analysed using mean-field Onsager reaction-field theory. This allowed us to evaluate the magnetic exchange interactions in $Ba_2MnTeO_6$ and $Ba_2MnWO_6$. We were able to describe the magnetic diffuse scattering above $T_N$ with a simple $J_1$-$J_2$ Heisenberg model. We found $J_1$ = -3.25(3) K and $J_2$ = 0.41(2) K for $Ba_2MnTeO_6$ and $J_1$ = -1.08(1) K and $J_2$ = -0.88(1) K $Ba_2MnWO_6$. The exchange constants are fully consistent with our previous linear spin wave theory analysis of the inelastic neutron scattering measured in the magnetically ordered state measured from the same samples [17,18]. Moreover, the fitting allowed us to estimate the ordering temperatures and to correctly predict the propagation vectors of the ordered phases. These results highlight the potential of Onsager reaction-field fitting for the determination of magnetic interactions in frustrated magnets.

**Acknowledgements**

The authors are grateful to Dr Joseph Paddison and Dr Andrew Wildes for fruitful discussions. Dr Paddison is also thanked for technical assistance with the SPINTERACT fitting. O.H.J.M., C.E.P., H.M.M. and E.J.C. acknowledge funding by the Leverhulme Trust Project Grant RPG-2017-109. O.H.J.M. is grateful for funding through Leverhulme Trust Early Career Fellowship ECF-2021-170. The magnetic property measurements were carried out at the Midlands Mag-Lab facility supported by the UK Engineering and Physical Sciences Research Council (EPSRC) grant EP/V028774/1. The authors are grateful for beamtime on the D7 diffuse scattering spectrometer at Institut Laue-Langevin under proposal number 5-32-909.

O.H.J.M., C.E.P., H.C.W. and E.J.C. conceived and planned the study. The samples were synthesised by C.E.P. and H.M.M. The magnetic susceptibility was measured by O.H.J.M. and R.T. The magnetic diffuse scattering was measured by O.H.J.M., C.E.P., L.M.-T., H.C.W. and E.J.C. The magnetic diffuse scattering was analysed by O.H.J.M., C.E.P. and R.T. O.H.J.M. wrote the manuscript with contributions from all authors.

**Magnetic diffuse scattering of the *S* = 5/2 fcc antiferromagnets Ba$_2$MnTeO$_6$ and Ba$_2$MnWO$_6$**

Otto H. J. Mustonen,[1,2]* Charlotte E. Pughe,[3] Lucile Mangin-Thro,[4] Robert Tarver,[2] Heather M. Mutch,[3] Helen C. Walker,[5] Edmund J. Cussen[6,3]*

[1]Department of Chemistry and Materials Science, Aalto University, FI-00076 Aalto, Finland

[2]School of Chemistry, University of Birmingham, Birmingham B15 2TT, United Kingdom

[3]Department of Material Science and Engineering, University of Sheffield, Mappin Street, Sheffield S1 3JD, United Kingdom

[4]Institut Laue-Langevin, 71 avenue des martyrs, 38000 Grenoble, France

[5]ISIS Pulsed Neutron and Muon Source, STFC Rutherford Appleton Laboratory, Harwell Campus, Didcot OX11 0QX, United Kingdom

[6]TU Dublin - School of Chemical and BioPharmaceutical Sciences, Grangegorman, D07 ADY7, Ireland

## Table of Contents

## SPINVERT fits of magnetic diffuse scattering

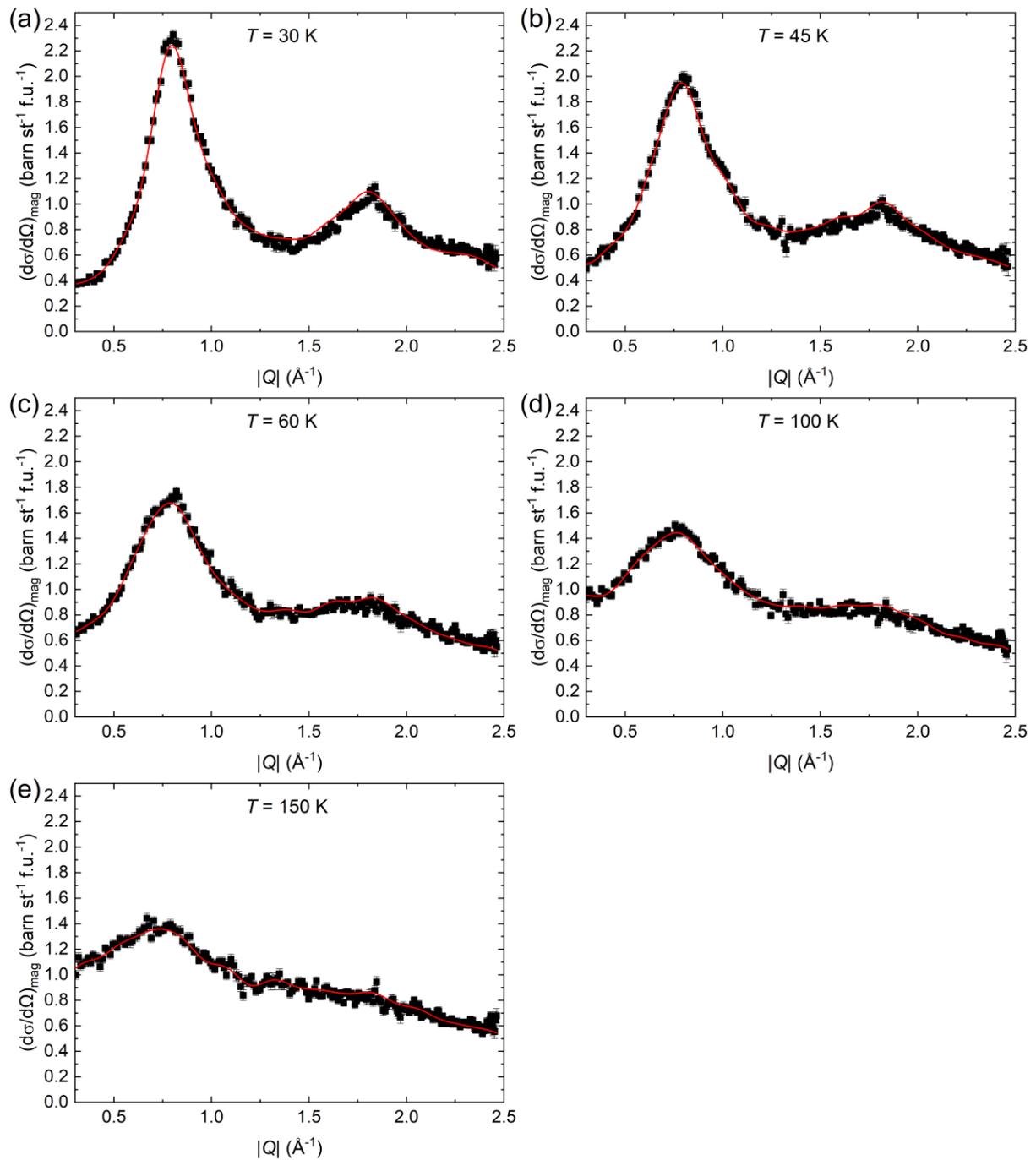

Figure S1. Magnetic diffuse scattering of $Ba_2MnTeO_6$ at various temperatures fitted using SPINVERT. (a) 30 K, (b) 45 K, (c) 60 K, (d) 100 K, (e) 150 K. The radial spin correlation functions obtained from these fits are shown in Figure 3(c) of the main paper.

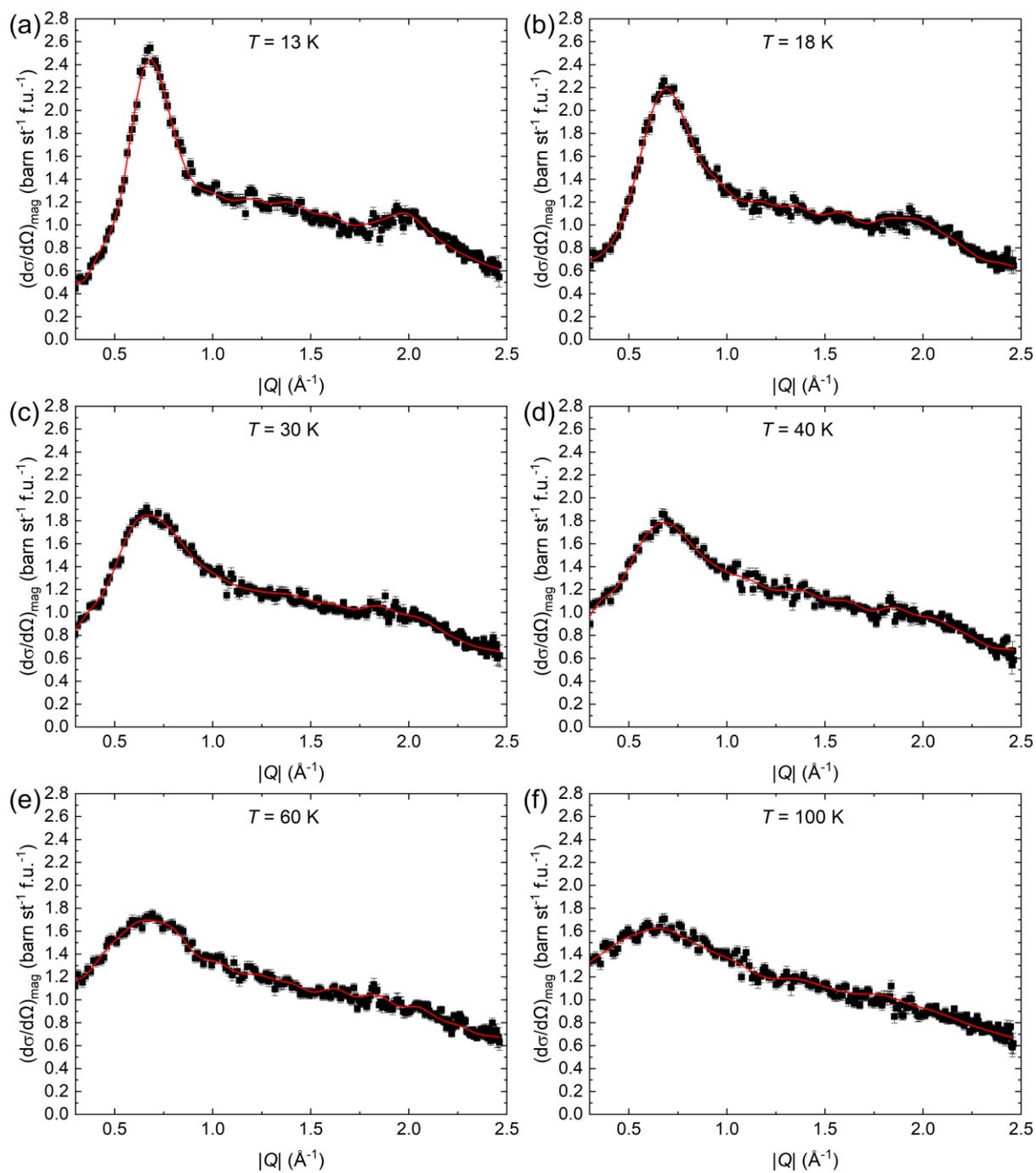

Figure S2. Magnetic diffuse scattering of $Ba_2MnWO_6$ at various temperatures fitted using SPINVERT. (a) 13 K, (b) 18 K, (c) 30 K, (d) 40 K, (e) 60 K, (f) 100 K. The radial spin correlation functions obtained from these fits are shown in Figure 3(d) of the main paper.

## SPINTERACT fits at a single temperature

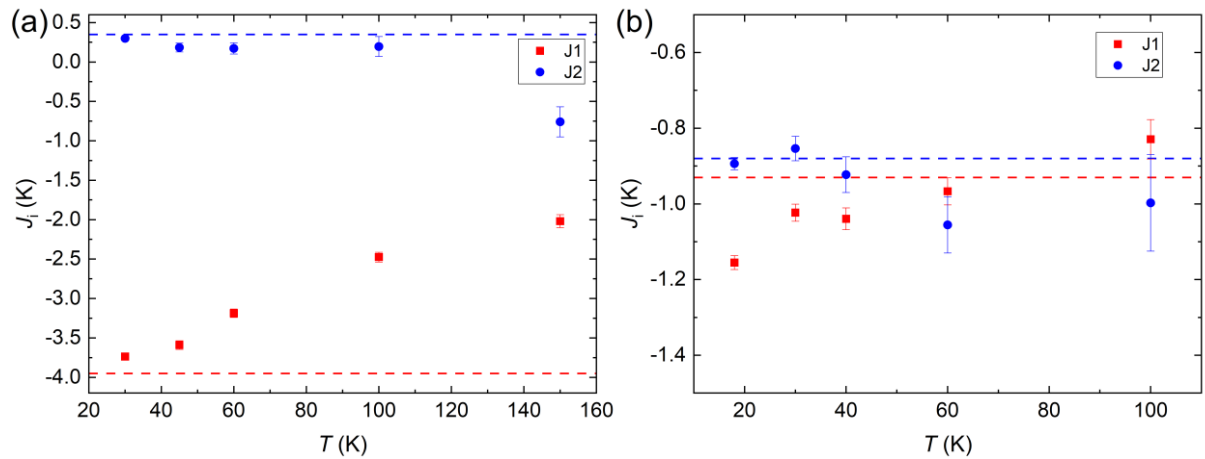

Figure S3. Magnetic interactions of (a) $Ba_2MnTeO_6$ and (b) $Ba_2MnWO_6$ obtained from single-temperature Onsager reaction-field fits. The dashed lines show the values obtained from inelastic neutron scattering [1,2]. For $Ba_2MnTeO_6$, the interactions are close to INS results only at low temperatures and start to deviate with increasing temperature. For $Ba_2MnWO_6$, the obtained parameters do not have a clear temperature dependence.